\documentclass[12pt]{article}
\usepackage[dvips]{graphicx}
\usepackage{amssymb,amsmath,amsfonts,palatino,amsthm,txfonts,accents}
\usepackage{amssymb,mathtools}
\usepackage[sans]{dsfont}
\usepackage[square, comma, sort&compress]{natbib}
\newcommand{\beq}{\begin{equation}}
\newcommand{\eeq}{\end{equation}}

\newcommand{\f}{\begin{equation}}
\newcommand{\ff}{\end{equation}}

\newtheorem{theorem}{Theorem}

\begin{document}

%%%%%%%%%%%%%%%%%%%%%%%%%%%%%%%%%%%%%%%%%%%%%%%%
\title{Effective Theory of Braid Excitations of Quantum Geometry in terms of Feynman Diagrams}
\author{Yidun Wan\thanks{Email address:
ywan@perimeterinstitute.ca}
\\
\\
\\
Perimeter Institute for Theoretical Physics,\\
31 Caroline st. N., Waterloo, Ontario N2L 2Y5, Canada, and \\
Department of Physics, University of Waterloo,\\
Waterloo, Ontario N2J 2W9, Canada\\}
\date{September 25, 2008}
\maketitle
\vfill
\begin{abstract}
We study interactions amongst topologically conserved excitations of
quantum theories of gravity, in particular the braid excitations of
four valent spin networks. These have been shown previously to
propagate and interact under evolution rules of spin foam models. We
show that the dynamics of these braid excitations can be described
by an effective theory based on Feynman diagrams. In this language,
braids which are actively interacting are analogous to bosons, in
that the topological conservation laws permit them to be singly
created and destroyed. Exchanges of these excitations give rise to
interactions between braids which are charged under the topological
conservation rules.
\end{abstract}
\vfill
\newpage
\tableofcontents
\newpage

\section{Introduction}

A substantial amount of work\cite{Bilson-Thompson2006, Hackett2007,
LouNumber, Wan2007, LeeWan2007, HackettWan2008, HeWan2008b,
HeWan2008a, Isabeau2008} has been done recently towards emergent
matter as topological invariants of framed spin networks, embedded
in a topological three-manifold\cite{Bilson-Thompson2006,
Hackett2007, LouNumber, Wan2007, LeeWan2007, HackettWan2008,
HeWan2008b, HeWan2008a} and unembedded\cite{Isabeau2008}, as
fundamental states of quantum gravity and quantum geometry. Some
papers study the case of framed three-valent spin
networks\cite{Bilson-Thompson2006, Hackett2007, LouNumber,
Isabeau2008}, which were motivated by the pioneer work of
Bilson-Thompson \cite{Bilson-Thompson2005}. Others investigate the
three-strand braids of four-valent spin networks embedded in a
topological three manifold\cite{LeeWan2007, HackettWan2008,
HeWan2008b, HeWan2008a}.

The aim of this paper is to present an effective description of
these emergent local excitations\cite{fotini-david,
Bilson-Thompson2006, LeeWan2007} in the language of Feynman
diagrams. As an analogy effective emergent degrees of freedom have
been realized in condensed matter physics. For example, there are
emergent quasiparticles as collective modes in some condensed matter
systems, such as phonons and rotons in superfluid He$^4$. Another
example is the string-net condensation which gives rise to low
energy effective gauge theories\cite{WenXiaogang}.

The three-valent case considered in \cite{Bilson-Thompson2006,
Hackett2007, LouNumber} is limited by lack of creation and
annihilation of the topological invariants\cite{Hackett2007} which
are considered corresponding to Standard Model
particles\cite{LouNumber}. This partly motivates our work in the
four-valent case\cite{Wan2007, LeeWan2007, HackettWan2008,
HeWan2008b, HeWan2008a} in which the topological excitations can
propagate and interact under the dual Pachner
moves\cite{LeeWan2007}. The four-valent spin networks here can be
understood as those naturally occur in spin foam
models\cite{spin-foam}, or in a more generic sense as the original
proposal of spin networks put forward by Penrose\cite{Penrose}.
Another motivation of working with four-valent spin networks is that
the vertices correspond to three-dimensional space\footnote{Note
that this correspondence is only between a 4-valent vertex and a 3d
simplicial complex. The precise correspondence with a (continuum)
physical 3d space remains a big open question.}.

The dynamical entities being studied in the four-valent case are
three-strand braids, each of which is formed by the three common
edges of two adjacent nodes of a spin network. With the graphic
notation and the classification of these three-strand braids
obtained in \cite{Wan2007}, \cite{LeeWan2007} discovers that stable
three-strand braids, which are thought to be local
excitations\cite{LeeWan2007} exist under certain stability
condition. Among the stable braids, there is a small class of
braids, called \textbf{propagating braids}, which can propagate on
the spin network. The braid propagation is chiral, in the sense that
some braids can only propagate to their left in their ambient local
graphs, whereas some only propagate to their right and some do
both\cite{Wan2007, LeeWan2007}. There is another small class of
braids, the \textbf{actively-interacting braids}; each is two-way
propagating and is able to merge with its neighboring braid to form
a new one\cite{LeeWan2007}. Stable braids that are not propagating
are christened \textbf{stationary braids}.

Keeping the graphic calculus of the braids developed in
\cite{Wan2007, LeeWan2007}, \cite{HackettWan2008, HeWan2008b}
proposed an algebraic calculus, which has been used to find out the
conserved quantities of braids, which are mapped to possible quantum
numbers such as electric charges, and the C, P, and T of our
braids\cite{HeWan2008b, HeWan2008a}.

In addition, \cite{HeWan2008b} shows that all actively interacting
braids form a noncommutative algebra of which the product (binary
operation) is braid interaction, and that an actively interacting
braid behaves like a map, taking a non-actively interacting braid to
another non-actively interacting one. However, each braid
interaction of the type discussed so far in \cite{LeeWan2007,
HackettWan2008, HeWan2008b, HeWan2008a} must always involve at least
one actively-interacting braid. In this paper we will investigate a
new type of interaction which takes two adjacent braids to another
two adjacent braids via exchanging a virtual actively interacting
braid. This seems to imply that actively interacting braids behave
like bosons whereas the others - in particular the chiral
propagating braids - may be candidates for fermions.

Most importantly, the dynamics of braids can be represented by braid
Feynman diagrams, based on which an effective field theory of the
braids exists. Our main results are summarized as follows.
\begin{enumerate}
\item Two neighboring braids may have an exchange interaction, i.e. they interact via exchanging a virtual actively interacting braid, resulting
in two different adjacent braids.
\item Effective twists and effective states of braids are conversed under exchange
interaction.
\item Exchange interaction is asymmetric in general; however,
conditions for an exchange interaction to be symmetric are given.
\item Braids can radiate actively interacting braids.
\item Actively interacting braids behave
analogously to bosons.
\item Braid Feynman diagrams representing braid dynamics are
proposed.
\item A constraint on probability amplitudes of braid dynamics is
obtained.
\end{enumerate}

In most of this paper we do not take into account the labels which
usually grace the edges and nodes of spin networks because the
existence and stability of the braid excitations we study do not
depend on them. Nevertheless, this paper will urge the incorporation
of spin network labels, in the traditional way or in a more
generalized way, in our next work, such that methods in certain spin
foam models, in group field theories, or in tensor categories can be
adopted (and modified).

\section{Notation and Division of Braids}
\begin{figure}
[h]
\begin{center}
\includegraphics[
height=2.1715in, width=2.8193in
]%
{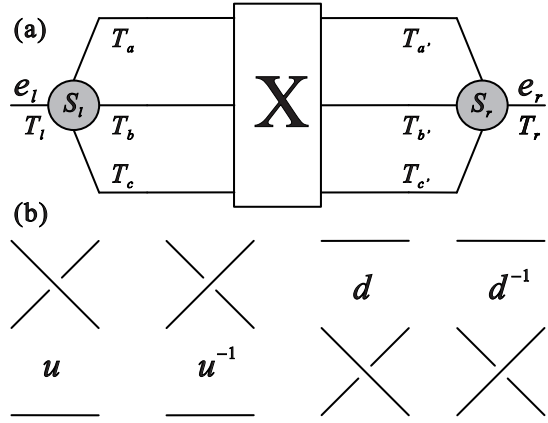}%
\caption{(a) is a generic 3-strand braid diagram. $S_l$ and $S_r$
are the states of the left and right \textbf{end-node}s
respectively, valued in $\{+,-\}$. $X$ represents a sequence of
crossings, from left to right, formed by the three strands. $(T_a,
T_b,T_c)$ is the triple of the \textbf{internal twists} respectively
on the three strands from top to bottom, on the left of $X$. $T_l$
and $T_r$, called \textbf{external twists}, are respectively on the
two external edges $e_l$ and $e_r$. All twists are valued in
$\mathbb{Z}$ in units of $\pi/3$\cite{Wan2007}. (b) shows the four
generators of $X$, which are also generators of the braid group
$B_3$.}%
\label{braid}%
\end{center}
\end{figure}

A 3-strand braid, formed by the three common edges of two adjacent
nodes, is in fact an equivalence class of diffeomorphic local sub
networks of the whole framed spin network embedded in a topological
3-manifold. We study a braid through its 2-dimensional projection,
called a \textbf{braid diagram}. We therefore will not distinguish
braids from braid diagrams unless an ambiguity arises. A generic
example of such a braid diagram is depicted in Fig. \ref{braid}(a).
Although the properties of a braid should not depend on which
representative is chosen for the equivalence class, in some cases an
appropriate representative makes things easier.

In \cite{LeeWan2007} our choice was to represent an equivalence
class by a braid diagram which has zero external twists. This
simplifies the interaction condition and the graphic calculus
developed in \cite{Wan2007, LeeWan2007}. Each class has one and only
one such representative. Thus a braid represented this way is said
to be in its \textbf{unique representation}. This representation is
important because according to \cite{HeWan2008b}, each multiplet of
actively interacting braids under the discrete transformations,
which are analogous to the C, P, and T in particle physics, found in
\cite{HeWan2008b} is uniquely characterized by a non-negative
integer, viz the number of crossings of the braids in their unique
representations in the multiplet. In this paper we will use the
unique representation exclusively.

Along with the graphic calculus, an algebraic notation and the
corresponding calculational method of braid interactions were put
forward in \cite{HackettWan2008, HeWan2008b}. This algebraic
formalism plays a key role in finding the conserved quantities of
braids and the discrete transformations mapped to C, P, T, and their
combinations, in \cite{HackettWan2008, HeWan2008b, HeWan2008a}. We
will use both the graphic and algebraic calculus in this paper. Thus
let us briefly review the algebraic notation.

The generic braid shown in Fig. \ref{braid}(a) is denoted by
$$\left._{T_l}^{S_l}\hspace{-0.5mm}[(T_a,T_b,T_c)\sigma_X]^{S_r}_{T_r}\right.,$$
where the crossing sequence $X$ is an element of the braid group
$B_3$, and the twists, $T_a$, $T_b$, etc, take values in
$\mathbb{Z}$. A crossing sequence is generated by the four
generators shown in Fig. \ref{braid}(b), and can then be written as
$X=x_1\dots x_i\dots x_n, 1\leq i\leq n\in\mathbb{N}$ with
$x_i\in\{u,d,u^{-1},d^{-1}\}$. It is useful to define the quantity
$X^{-1}=x^{-1}_n\dots x^{-1}_i\dots x^{-1}_1$ for
$X$\cite{HeWan2008a}. So clearly $XX^{-1}=X^{-1}X=\mathbb{I}$,
meaning no crossing. The generators are assigned integral values
according to their handedness, namely $u=d=1$ and
$u^{-1}=d^{-1}=-1$. Therefore, crossings of a braid can also be
summed over to obtain an integer, the \textbf{crossing number}:
$\sum_{i=1}^{|X|}x_i$, of the braid, where $|X|$ is the number of
crossings.

The $X$ of a braid induces a permutation $\sigma_X$, an element in
the permutation group $S_3$, of the three strands of the braid. The
triple of internal twists on the left of $X$ and the one of the
right of $X$ are thus related by
$(T_a,T_b,T_c)\sigma_X=(T_{a'},T_{b'},T_{c'})$ and
$(T_a,T_b,T_c)=\sigma^{-1}_X(T_{a'},T_{b'},T_{c'})$. That is,
$\sigma_X$ is a left-acting function of the triple of internal
twists, while its inverse, $\sigma^{-1}_X$ is right-acting. The
inverse relation between $\sigma_X$ and $\sigma^{-1}_X$ is
understood as:
\begin{equation}
\sigma^{-1}_X\left((T_a,T_b,T_c)\sigma_X\right)=\left(\sigma^{-1}_X(T_a,T_b,T_c)\right)\sigma_X\equiv(T_a,T_b,T_c)
\label{eqSigmaTsigma}
\end{equation}
Note that the twists such as $T_a$ and $T_{a'}$ are abstract and
have no concrete meaning until their values and positions in a
triple are fixed. Thus, $(T_a,T_b,T_c)=(T'_a,T'_b,T'_c)$ means
$T_a=T'_a$, etc, and $-(T_a,T_b,T_c)=(-T_a,-T_b,-T_c)$. Note also
that the $\sigma_X$ in the out most layer in the notation should be
kept formal rather than an explicit element in $S_3$ because the $X$
is needed to record the crossing sequence. For example, if we obtain
a braid, say
$\left._{T_l}^{S_l}\hspace{-0.5mm}[((T_a,T_b,T_c)+\sigma^{-1}_d)\sigma_{ud}]^{S_r}_{T_r}\right.,$
in some process, then while we can write the $\sigma^{-1}_d$ as
$(2,\ 3)$ to complete the addition in the parenthesis, we should
keep $\sigma_{ud}$ as it is but not explicitly as $(2\ 3\ 1)$. The
relation between $\sigma_X$ and its inverse suggests that a generic
braid can be equally denoted by
$$\left._{T_l}^{S_l}\hspace{-0.5mm}[\sigma^{-1}_X(T_{a'},T_{b'},T_{c'})]^{S_r}_{T_r}\right..$$

Given this notation, we express a braid in its unique representation
as
$$\left._{\hspace{1mm}0}^{S_l}\hspace{-0.5mm}[(T_a,T_b,T_c)\sigma_X]^{S_r}_{0}\right.,$$
or simply
$\left.^{S_l}\hspace{-0.5mm}[(T_a,T_b,T_c)\sigma_X]^{S_r}\right.$,
dropping out the zero's.

We have shown the algebraic structure of the set of all stable
3-strand braids in \cite{HeWan2008b} under the braid interaction
found in \cite{LeeWan2007}. Hence, we choose to denote the set of
stable braids by $\mathfrak{B}^S_0$. However, for reasons which will
become transparent in the sequel we should enlarge
$\mathfrak{B}^S_0$ by adding two more braids:
\begin{equation}
B_0^{\pm}=\left._0^{\pm}\hspace{-0.5mm}[0,0,0]^{\pm}_0\right.,\label{eqB0pm}
\end{equation}
which are completely trivial\footnote{It is highly important to note
again that at this point spin network labels have not been taken
into account yet. When spin network labels are rather considered
there are two immediate consequences. 1) $B_0^{\pm}$ are not just
two braids, but infinite number of braids because they can be
colored by different sets of spin network labels, so is any other
braid in $\mathfrak{B}^S$. 2) $B_0^{\pm}$ are only trivial
topologically but not algebraically and neither physically.}.
$B_0^{\pm}$ are actually unstable according to \cite{LeeWan2007}
because they are dual to a 3-ball. If we tolerate their instability
they are obviously actively interacting. Bearing this in mind, let
us still call the enlarged set the set of stable braids but use
$\mathfrak{B}^S$ as its notation. According to \cite{LeeWan2007},
$\mathfrak{B}^S$ can be divided into the disjoint union of three
subsets: the subset of all actively interacting braids (including
$B_0^{\pm}$), the subset of all propagating braids which are do not
actively interact, and the subset of stationary braids which are
passive in all interactions and do not propagate; they are denoted
respectively by $\mathfrak{B}^b$, $\mathfrak{B}^f$, and
$\mathfrak{B}^s$. Therefore, we have
\begin{equation}
\mathfrak{B}^S=\mathfrak{B}^b\sqcup\mathfrak{B}^f\sqcup\mathfrak{B}^s.%
\label{eqDUnionB}
\end{equation}
Moreover, the set of propagating-only braids $\mathfrak{B}^f$ can be
further divided as
\begin{equation}
\mathfrak{B}^f=\mathfrak{B}^f_L\sqcup\mathfrak{B}^f_R\sqcup\mathfrak{B}^f_T,%
\label{eqDUnionBf}
\end{equation}
where $\mathfrak{B}^f_L$ denotes the subset of braids which only
propagate to the left, $\mathfrak{B}^f_R$ is the subset of braids
which only propagate to the right, and $\mathfrak{B}^f_T$ contains
two-way propagating braids.

Because a new type of interaction will be introduced in this paper,
to get rid of any possible confusion we call the interaction
introduced in \cite{LeeWan2007} the \textit{direct interaction} and
notate it by $+_{\mathrm{d}}$, while name the new interaction
\textit{exchange interaction} for reasons will be clear when it is
defined, denoted as $+_{\mathrm{e}}$. As a consequence, a bare $+$
sign between two braids only means the adjacency of the braids. As
pointed out in \cite{HeWan2008b}, $\mathfrak{B}^b$ is closed under
the direct interaction, i.e. $\mathfrak{B}^b
+_{\mathrm{d}}\mathfrak{B}^b \subseteq \mathfrak{B}^b$.

This will help to make the algebraic structure of braids and the
contact of our model with particle physics clearer.

\section{Exchange interaction}
\cite{LeeWan2007} has shown that two neighboring braids may interact
and merge into a new braid. This process is called direct
interaction. One of the two braids in a direct interaction must be
an actively interacting braid. That is, we only have the following
possible direct interactions: $\mathfrak{B}^b +_{\mathrm{d}}
\mathfrak{B}^b$, $\mathfrak{B}^b +_{\mathrm{d}}
\mathfrak{B}\setminus\mathfrak{B}^b$. Fortunately, there exists, as
we are to show, another type of interaction, namely the exchange
interaction, which can take two adjacent braids in
$\mathfrak{B}\setminus\mathfrak{B}^b$ to another two neighboring
ones in the same set. In fact, exchange interaction can be defined
on the whole $\mathfrak{B}^S$ as a map, $+_{\mathrm{e}}:
\mathfrak{B}^S\times\mathfrak{B}^S\rightarrow\mathfrak{B}^S\times\mathfrak{B}^S$.
It will be clear shortly that there is always an exchange of a
virtual actively interacting braid in an exchange interaction,
giving why exchange interaction is so christened. It is useful to
keep track of the direction of the exchange of the actively
interacting braid during an exchange interaction. Therefore, we
differentiate a \textit{left} and a \textit{right exchange
interaction}, respectively denoted by
$\accentset{\leftarrow}{+}_{\mathrm{e}}$ and
$\accentset{\rightarrow}{+}_{\mathrm{e}}$. The arrow indicates the
"flow" of the virtual actively interacting braid.

The graphic definition of right exchange interaction is illustrated
in Fig. \ref{type2IntDef}. The left one can be defined likewise. We
now explain with the help of our algebraic notation the process in
detail as follows.
\begin{enumerate}
\item We begin with the two adjacent braids,
$B_1,B_2\in\mathfrak{B}^S$ in Fig. \ref{type2IntDef}(a). Their
algebraic forms are:
$B_1=\left.^{S_{1l}}\hspace{-0.5mm}[(T_{1a},T_{1b},T_{1c})\sigma_{X_{1A}X_{1B}}]^{S}\right.$
and
$B_2=\left.^{S}\hspace{-0.5mm}[(T_{2a},T_{2b},T_{2c})\sigma_{X_2}]^{S_{2r}}\right.$.
$B_1$'s right end-node and $B_2$'s left end-node are set in the same
state, $S$, to satisfy the interaction condition\cite{LeeWan2007}.
We also assume that $B_1$ is right-reducible (not necessarily fully
right-reducible), such that the crossing sequence of $B_1$ has a
maximal reducible segment, say $X_{1B}$, as shown in the figure. The
reason for this will be clear soon. (Similarly, for left exchange
interaction one should have $B_2$ left-reducible. If $B_1$ is
right-reducible and $B_2$ is left-reducible, both left and right
exchange interactions may occur but they lead to different results
in general.)
\item A $2\rightarrow 3$ move on the two nodes in the state $S$ now leads to Fig.
\ref{type2IntDef}(b). The dashed lines simply means the crossing
relation between the green lines and black ones can not be
determined unless we know what exactly the state $S$ is.
\item Because $X_{1B}$ is the reducible part of the crossing
sequence of $B_1$, according to \cite{LeeWan2007} one can translate
the three new nodes - one in state $S$ and two in $-S$ - in (b)
along with the edges $g, g'$, and $g''$ to the left of $X_{1B}$, and
rearrange them by equivalence moves defined in \cite{Wan2007} in a
proper configuration ready for a $3\rightarrow 2$ move, as seen in
Fig. \ref{type2IntDef}(c). This is why we assume $B_1$ is
\begin{figure}[hp]
\begin{center}
  \includegraphics[height=4.91in,width=2.92in]{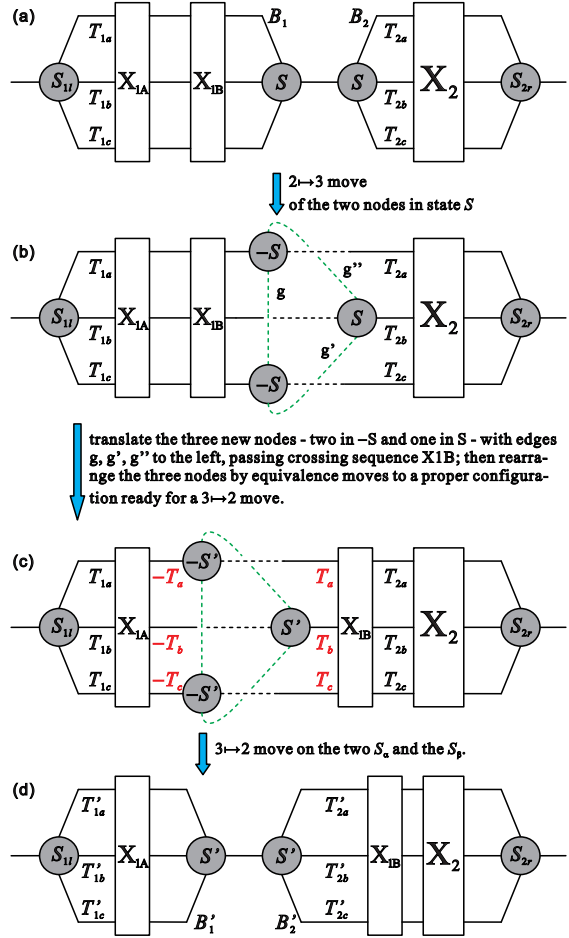}
  \caption{Definition of right exchange interaction:
  $B_1\overset{\rightarrow}{+}_{\mathrm{e}}B_2\rightarrow B'_1+B'_2$,
  $B_1,B_2,B'_1,B'_2\in\mathfrak{B}^S$. In (b), the dashed lines
  emphasize the dependence of the $2\rightarrow 3$
  move on the state $S$ of the two nodes on which the move is taken.
  In (c) and (d), $S'=(-)^{|X_{1B}|}S$. In (c), the dashed line
  means that the configuration depends on $S'$. In (d), $(T'_{1a},T'_{1b},T'_{1c})$ and $(T'_{2a},T'_{2b},T'_{2c})$ are
  the internal twists of $B'_1$ and $B'_2$ respectively, which are explained in the
  text.}
  \label{type2IntDef}
\end{center}
\end{figure}
right-reducible; otherwise, the translation is not viable. Due to
\cite{LeeWan2007}, in Fig. \ref{type2IntDef}(c) the three nodes are
shuffled by the translation and their states are related to $S$ by
$S'=(-)^{|X_{1B}|}S$. However, because $S$ is arbitrary dashed lines
are also used in Fig. \ref{type2IntDef}(c) for undetermined crossing
relation between the green edges and the black ones. This procedure
introduces twists in pair on the strands: $-(T_a,T_b,T_c)$ and
$(T_a,T_b,T_c)$, labeled in red in Fig. \ref{type2IntDef}(c).

Note that the triples $(T_{1a},T_{1b},T_{1c})$ and $-(T_a,T_b,T_c)$
are not existing on the strands separately. One should add
$-(T_a,T_b,T_c)$ to $(T_{1a},T_{1b},T_{1c})$ with the permutation
induced by $X_{1A}$ taken into account, i.e.
$(T_{1a},T_{1b},T_{1c})-\sigma^{-1}_{X_{1A}}(T_a,T_b,T_c)$. The
relation between $(T_{2a},T_{2b},T_{2c})$ and $(T_a,T_b,T_c)$ is
likewise. This cannot be represented in the figure, which is a
limitation of the graphic notation.
\item We then perform the $3\rightarrow 2$ move and arrive at Fig.
\ref{type2IntDef} (c) which shows two new adjacent braids, $B'_1$
and $B'_2$, related to $B_1$ and $B_2$ by
\begin{align}
(T'_{1a},T'_{1b},T'_{1c})&=(T_{1a},T_{1b},T_{1c})-\sigma^{-1}_{X_{1A}}(T_a,T_b,T_c)\label{eqR1}\\
(T'_{2a},T'_{2b},T'_{2c})&=(T_a,T_b,T_c)+\sigma^{-1}_{X_{1B}}(T_{2a},T_{2b},T_{2c}).\label{eqR2}
\end{align}
This completes the right exchange interaction,
$B_1\accentset{\rightarrow}{+}_{\mathrm{e}}B_2\rightarrow
B'_1+B'_2$.
\end{enumerate}
\begin{figure}[h]
\begin{center}
  \includegraphics[height=1.12in,width=2.21in]{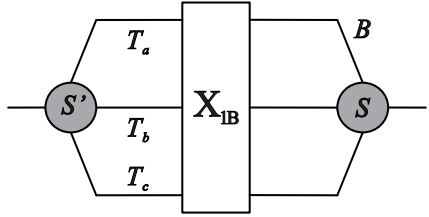}
  \caption{The actively interacting braid exchanged from $B_1$ to $B_2$, during their exchange interaction, in Fig.
  \ref{type2IntDef}.}
  \label{Bexchange}
\end{center}
\end{figure}

It can be shown, according to \cite{LeeWan2007, HeWan2008b}, that
the only possible triple $(T_a,T_b,T_c)$ in Fig.
\ref{type2IntDef}(c) is exactly the same as the triple of internal
twists of the actively interacting braid,
$B=\left.^{S'}\hspace{-0.5mm}[(T_a,T_b,T_c)\sigma_{X_{1B}}]^{S}\right.$,
in Fig. \ref{Bexchange}. Note that as proved in
\cite{HackettWan2008}, for an actively interacting braid of the form
in Fig. \ref{Bexchange}, $S'=(-)^{|X_{1B}|}S$. Hence, $B$'s left and
right end-nodes are respectively in the same states as that of the
left end-node of $B'_2$ in Fig. \ref{type2IntDef}(d) and that of the
right end-node of $B_1$ in Fig. \ref{type2IntDef}(a). Thus the form
of braid $B'_2$ in Fig. \ref{type2IntDef}(d) due to the final
$3\rightarrow 2$ move in the interacting process must be precisely
the result of the direct interaction of $B$ and $B_2$. That is, by
\cite{HeWan2008b} we have
\begin{align*}
B +_{\mathrm{d}} B_2 &=
\left.^{S'}\hspace{-0.5mm}[(T_a,T_b,T_c)\sigma_X]^{S}\right.
+_{\mathrm{d}}
\left.^{S}\hspace{-0.5mm}[(T_{2a},T_{2b},T_{2c})\sigma_{X_2}]^{S_{2r}}\right.\\
&\rightarrow
\left.^{S'}\hspace{-0.5mm}[((T_a,T_b,T_c)+\sigma^{-1}_{X_{1B}}(T_{2a},T_{2b},T_{2c}))\sigma_{X_2}]^{S_{2r}}\right.=B'_2,
\end{align*}
which validates the relation in Eq. \ref{eqR2}.

Therefore, the process of the right exchange interaction defined by
Fig. \ref{type2IntDef} is like that $B_1$ gives out the actively
interacting braid $B$ in Fig. \ref{Bexchange} which is then combined
with $B_2$ by a direct interaction. In other words, $B_1$ and $B_2$
interact with each other by exchanging a virtual actively
interacting braid $B$, and become $B'_1$ and $B'_2$. Or one may say
that an exchange interaction is mediated by an actively interacting
braid. This implies an analogy between actively interacting braids
and bosons\footnote{More generally, this should imply the analogy
between actively interacting braids and particles that mediate
interactions, which should potentially include super partners of
gauge bosons.}. Note that in an exchange interaction, there does not
exist an intermediate state in which only the virtual actively
interacting braid is present because our definition of a braid
requires the presence of its two end-nodes.

The above can be summarized by the following theorem as the first
main result of the paper. (The case of left exchange interaction is
similar.)
\begin{theorem}\label{theoType2Int}
Given two adjacent braids, $B_1,B_2\in\mathfrak{B}^S$,
$B_1=\left.^{S_{1l}}\hspace{-0.5mm}[(T_{1a},T_{1b},T_{1c})\sigma_{X_{1A}X_{1B}}]^{S}\right.$
is on the left and is right-reducible with $X_{1B}$ the reducible
segment of its crossing sequence, and
$B_2=\left.^{S}\hspace{-0.5mm}[(T_{2a},T_{2b},T_{2c})\sigma_{X_2}]^{S_{2r}}\right.$,
there exists a braid $B\in\mathfrak{B}^b$,
$B=\left.^{S'}\hspace{-0.5mm}[(T_a,T_b,T_c)\sigma_{X_{1B}}]^{S}\right.$
with $S'=(-)^{|X_{1B}|}S$, such that it mediates the exchange
interaction of $B_1$ and $B_2$ to create $B'_1,
B'_2\in\mathfrak{B}^S$, i.e.
\begin{equation}
\begin{aligned}
& B_1\accentset{\rightarrow}{+}_{\mathrm{e}}B_2\\
&\rightarrow
\left.^{S_{1l}}\hspace{-0.5mm}[((T_{1a},T_{1b},T_{1c})-\sigma^{-1}_{X_{1A}}(T_a,T_b,T_c))\sigma_{X_{1A}}]^{S'}\right.
+
\left.^{S'}\hspace{-0.5mm}[((T_a,T_b,T_c)+\sigma^{-1}_{X_{1B}}(T_{2a},T_{2b},T_{2c}))\sigma_{X_2}]^{S_{2r}}\right.\\
&= B'_1+B'_2.
\end{aligned}\label{eqRtype2Int}
\end{equation}
\end{theorem}

It is important to remark that the reducibility of either braid in
an exchange interaction is not necessary if we have included braid
$B_0^{\pm}$ in Eq. \ref{eqB0pm}. For example, for two neighboring
irreducible braids one can still have the steps in Fig.
\ref{type2IntDef}(a) and (b), then skip over Fig.
\ref{type2IntDef}(c) because there are no reducible crossing segment
to be translated through, and directly take the step in Fig.
\ref{type2IntDef}(d). Such a procedure is still dynamical because of
the action of evolution moves and thus can be considered a special
case of exchange interaction. Needless to say, the actively
interacting braid being exchanged in such an interaction is one of
the two trivial braids $B_0^{\pm}$. This ensures that exchange
interaction is a map on $\mathfrak{B}^S\times\mathfrak{B}^S$.

Another important remark is that two braids can have exchange
interactions in different ways, in contrary to direct interaction.
The occurrence of an exchange interaction on two braids does not
have to exhaust the maximal reducible crossing segment of the
reducible braid. For example as in Fig. \ref{type2IntDef}, since
$X_{1B}$ is the maximal reducible crossing segment, we may take it
to be the concatenation of two reducible crossing segments, i.e.
$X_{1B}=X'_{1B}X_{1B}''$, then the translation taking Fig.
\ref{type2IntDef}(b) to (c) is allowed to terminate after passing
through $X_{1B}''$ which becomes the crossing sequence of the
virtual actively interacting braid in this new process. This
certainly leads to two braids different from those in Fig.
\ref{type2IntDef}(d). In an ideal scenario, each possible way of the
exchange interaction of two braids should have certain probability
to occur.

There is an analogy of this in particle physics. Since quarks have
both electric and color charges, both photons and gluons can mediate
forces on quarks. However, the relation between actively interacting
braids and bosons is yet not an actual identification. In fact, if
each actively interacting braid corresponded to a boson, there would
be too many "bosons". The underlining physics of that two braids can
have exchange interactions in different ways deserves further
studies.

It should be emphasized that each individual exchange interaction is
a process giving rise to a unique result\footnote{With spin network
labels the result is not unique any more because two topologically
equal braids can be decorated by different sets of labels, and an
interaction should result in a superposition of braids labeled
differently.}. For preciseness, an expression like
$B_1\accentset{\rightarrow}{+}_{\mathrm{e}}B_2$ is only formal. Only
when the exact forms of $B_1$ and $B_2$, in which their reducible
segments are explicitly present, are given,
$B_1\accentset{\rightarrow}{+}_{\mathrm{e}}B_2$ acquires a precise
and unique meaning. In computing an exchange interaction, we have to
specify explicitly our choice of the reducible crossing segment of
the braid which gives out the virtual actively interacting braid
which depends on this choice. For any such choice Theorem
\ref{theoType2Int} holds.

\cite{LeeWan2007, HackettWan2008, HeWan2008b} have shown that there
are two representative-independent conserved quantities of stable
braids, namely the effective twist $\Theta$ and the effective state
$\chi$, the former of which is additively conserved under direct
interaction while the latter is multiplicatively conserved. Due to
the representative-independence of these two quantities, we can
write down their expressions for an arbitrary braid
$B\in\mathfrak{B}^S$ in its unique representation, say
$B=\left.^{S_l}\hspace{-0.5mm}[(T_a,T_b,T_c)\sigma_X]^{S_r}\right.$,
as $\Theta_B=\sum\limits^c_{i=a}T_i - 2\sum\limits^{|X|}_{j=1}x_j$
and $\chi_B=(-)^{|X|}S_lS_r$. These two quantities are important.
\cite{HeWan2008a} has shown that $\Theta$ or certain function of
$\Theta$ of a braid can be accounted for the "electric" charge of
the braid. That is, braids can be charged. On the other hand, for
actively interacting braids $\chi$ is a characteristic quantity
because $\chi\equiv 1$ for any such braid. The following theorem
presents that exchange interaction also preserves these two
quantities in the same way as direct interaction does.
\begin{theorem}
Given two neighboring stable braids, $B_1,B_2\in\mathfrak{B}^S$,
such that an exchange interaction (left or right or both) on them is
doable, i.e. $B_1+_{\mathrm{e}}B_2\rightarrow B'_1+B'_2$,
$B'_1,B'_2\in\mathfrak{B}^S$, the effective twist $\Theta$ is an
additive conserved quantity, while the effective state $\chi$ is a
multiplicative conserved quantity, namely
\begin{equation}
\begin{aligned}
\Theta_{B'_1}+\Theta_{B'_2} &\overset{+_{\mathrm{e}}}{=} \Theta_{B_1}+\Theta_{B_2}\\
\chi_{B'_1}\chi_{B'_2} &\overset{+_{\mathrm{e}}}{=}
\chi_{B_1}\chi_{B_2}
\end{aligned}\label{eqConserve}
\end{equation}
This conservation law is independent of the virtual actively
interacting braid being exchanged during the exchange interaction.
\end{theorem}
\begin{proof}
It is sufficient to prove this in the case of right exchange
interaction, the other cases follow similarly. One can assume $B_1$
and $B_2$ are in the form as they are in Theorem \ref{theoType2Int}.
Hence, according to Theorem \ref{theoType2Int} one can readily write
down
$$\Theta_{B_1}=\sum\limits^c_{i=a}T_{1i} -
2\sum\limits^{|X_1|}_{j=1}x_{1j},\ \
\Theta_{B_2}=\sum\limits^c_{i=a}T_{2i} -
2\sum\limits^{|X_2|}_{k=1}x_{2k},\ \ \chi_{B_1}=(-)^{|X_1|}S_{1l}S,
\ \ \chi_{B_2}=(-)^{|X_2|}SS_{2r},$$ where $X_1=X_{1A}X_{1B}$, and
\begin{align*}
&\Theta_{B'_1}=\sum\limits^c_{i=a}(T_{1i} - T_i)-
2\sum\limits^{|X_{1A}|}_{j=1}x_{1Aj},\ \ \chi_{B'_1}=(-)^{|X_{1A}|}S_{1l}S',\\
&\Theta_{B'_2}=\sum\limits^c_{i=a}(T_i+T_{2i}) -
2\sum\limits^{|X_{1B}|}_{m=1}x_{1Bm}-
2\sum\limits^{|X_2|}_{k=1}x_{2k},\ \
\chi_{B'_2}=(-)^{|X_{1B}X_2|}S'S_{2r},
\end{align*}
Hence we have the following:
\begin{align*}
\Theta_{B'_1}+\Theta_{B'_2} &= \sum\limits^c_{i=a}(T_{1i} - T_i)-
2\sum\limits^{|X_{1A}|}_{j=1}x_{1Aj} +
\sum\limits^c_{i=a}(T_i+T_{2i}) -
2\sum\limits^{|X_{1B}|}_{m=1}x_{1Bm}- 2\sum\limits^{|X_2|}_{k=1}x_{2k}\\
&= \sum\limits^c_{i=a}(T_{1i} + T_{2i}) -
2\sum\limits^{|X_1|}_{j=1}x_{1j}-2\sum\limits^{|X_2|}_{k=1}x_{2k}\\
&= \Theta_{B_1}+\Theta_{B_2},
\end{align*}
$$\chi_{B'_1}\chi_{B'_2}=
(-)^{|X_{1A}|}S_{1l}S'(-)^{|X_{1B}X_2|}S'S_{2r}=(-)^{|X_{1A}X_{1B}|}S_{1l}(-)^{|X_2|}S_{2r}=\chi_{B_1}\chi_{B_2},$$
where the use of
$\sum\limits^{|X_{1A}|}_{j=1}x_{1Aj}+\sum\limits^{|X_{1B}|}_{m=1}x_{1Bm}=\sum\limits^{|X_1|}_{j=1}x_{1j}$,
$S'^2=S^2\equiv +$, and $X_{1A}X_{1B}=X$ has been made.
\end{proof}

Therefore, exchanges of actively interacting braids give rise to
interactions between braids which are charged under the topological
conservation rules. The conservation of $\Theta$ is analogous to the
charge conservation in particle physics.

\subsection{Asymmetry of exchange interaction}
As in the case of direct interaction\cite{HeWan2008b}, exchange
interaction is not symmetric either. The asymmetry of exchange
interaction is two-fold. On the one hand, given
$B_1,B_2\in\mathfrak{B}^S$, which are both right-reducible
(left-reducible), in general
$B_1\accentset{\rightarrow}{+}_{\mathrm{e}} B_2\neq
B_2\accentset{\rightarrow}{+}_{\mathrm{e}} B_1$
($B_1\accentset{\leftarrow}{+}_{\mathrm{e}} B_2\neq
B_2\accentset{\leftarrow}{+}_{\mathrm{e}} B_1$). This is henceforth
called the \textit{asymmetry of the first kind}. On the other hand,
if $B_1$ and $B_2$ are respectively right- and left-reducible, then
in general $B_1\accentset{\rightarrow}{+}_{\mathrm{e}} B_2\neq
B_1\accentset{\leftarrow}{+}_{\mathrm{e}} B_2$, which is christened
the \textit{asymmetry of the second kind}. Most probably, the
interactions on either side of these inequalities simply does not
occur due to the interaction condition. Even when these interactions
do occur, the inequalities hold in general. However, we now show
that there are cases where exchange interaction can be symmetric.

An issue is that two braids may have several different exchange
interactions. It is then impossible for, say
$B_1\accentset{\rightarrow}{+}_{\mathrm{e}} B_2$, to be equal to
$B_2\accentset{\rightarrow}{+}_{\mathrm{e}} B_1$ for all possible
ways of how $B_1$ and $B_2$ may interact. The only precise question
we can answer is actually, taking right exchange interaction as an
example: For any $B_1$ and $B_2$, among all possible ways of
$B_1\accentset{\rightarrow}{+}_{\mathrm{e}} B_2$ and
$B_2\accentset{\rightarrow}{+}_{\mathrm{e}} B_1$, does there exist a
way in which the reducible crossing segments of $B_1$ and $B_2$ are
chosen, such that in this way
$B_1\accentset{\rightarrow}{+}_{\mathrm{e}} B_2=
B_2\accentset{\rightarrow}{+}_{\mathrm{e}} B_1$?

We first study the the asymmetry of the first kind. It suffices to
check the right exchange interaction and the left one follows
likewise. We assume
$B_1=\left.^{S_{1l}}\hspace{-0.5mm}[(T_{1a},T_{1b},T_{1c})\sigma_{X_1X'_1}]^{S_{1r}}\right.$
and
$B_2=\left.^{S_{2l}}\hspace{-0.5mm}[(T_{2a},T_{2b},T_{2c})\sigma_{X_2X'_2}]^{S_{2r}}\right.$,
with $X'_1$ and $X'_2$ certain choices of the reducible crossing
segments of respectively $B_1$ and $B_2$. The fact that both
$B_1\accentset{\rightarrow}{+}_{\mathrm{e}} B_2$ and
$B_2\accentset{\rightarrow}{+}_{\mathrm{e}} B_1$ are assumed to be
legal requires that $S_{1r}=S_{2l}=S$ and $S_{2r}=S_{1l}=S'$ for
some $S$ and $S'$. Hence, we have
$B_1=\left.^{S'}\hspace{-0.5mm}[(T_{1a},T_{1b},T_{1c})\sigma_{X_1X'_1}]^{S}\right.$
and
$B_2=\left.^{S}\hspace{-0.5mm}[(T_{2a},T_{2b},T_{2c})\sigma_{X_2X'_2}]^{S'}\right.$.
With this, Theorem \ref{theoType2Int} immediately gives us
\begin{align*}
&B_1\accentset{\rightarrow}{+}_{\mathrm{e}} B_2\\
&\rightarrow
\left.^{S'}\hspace{-0.5mm}[((T_{1a},T_{1b},T_{1c})-\sigma^{-1}_{X_1}(T_a,T_b,T_c))\sigma_{X_1}]^{S''}\right.
+
\left.^{S''}\hspace{-0.5mm}[((T_a,T_b,T_c)+\sigma^{-1}_{X'_1}(T_{2a},T_{2b},T_{2c}))\sigma_{X'_1X_2X'_2}]^{S'}\right.
\end{align*}
\begin{align*}
&B_2\accentset{\rightarrow}{+}_{\mathrm{e}} B_1\\
&\rightarrow
\left.^{S}\hspace{-0.5mm}[((T_{2a},T_{2b},T_{2c})-\sigma^{-1}_{X_2}(T'_a,T'_b,T'_c))\sigma_{X_2}]^{S'''}\right.
+
\left.^{S'''}\hspace{-0.5mm}[((T'_a,T'_b,T'_c)+\sigma^{-1}_{X'_2}(T_{1a},T_{1b},T_{1c}))\sigma_{X'_2X_1X'_1}]^{S}\right.,
\end{align*}
where $S''=(-)^{|X'_1|}S$ and $S'''=(-)^{|X'_2|}S'$; $(T_a,T_b,T_c)$
and $(T'_a,T'_b,T'_c)$ are the triples of internal twists of the two
virtual actively interacting braids being exchanged in the two
interactions respectively. Requiring that the RHS of the two
interaction equations are equal term by term gives rise to
\begin{equation}
\left\{ %
\begin{array}
[c]{ll}%
S=S',&\ \ S''=S''' \medskip\\
X_1=X_2,&\ \ X'_1X_2X'_2=X'_2X_1X'_1
\end{array}
\right.,\label{eqSX}
\end{equation}
and
\begin{equation}
\left\{ %
\begin{array}
[c]{l}%
(T_{1a},T_{1b},T_{1c})-\sigma^{-1}_{X_1}(T_a,T_b,T_c)= (T_{2a},T_{2b},T_{2c})-\sigma^{-1}_{X_2}(T'_a,T'_b,T'_c)\medskip\\
(T_a,T_b,T_c)+\sigma^{-1}_{X'_1}(T_{2a},T_{2b},T_{2c})=(T'_a,T'_b,T'_c)+\sigma^{-1}_{X'_2}(T_{1a},T_{1b},T_{1c})
\end{array}
\right..\label{eqTtriple}
\end{equation}
Eq. \ref{eqSX} also implies $X'_1=X'_2$. We can then rewrite $S'$ as
$S$, $S'''$ as $S''$, and both $X_1X'_1$ and $X_2X'_2$ as $XX'$. As
a result, the two triples $(T_a,T_b,T_c)$ and $(T'_a,T'_b,T'_c)$
must be equal, which is understood by recalling the steps in Fig.
\ref{type2IntDef}. Thus Eq. \ref{eqTtriple} becomes
\begin{equation*}
\left\{ %
\begin{array}
[c]{l}%
(T_{1a},T_{1b},T_{1c})-\sigma^{-1}_{X}(T_a,T_b,T_c)= (T_{2a},T_{2b},T_{2c})-\sigma^{-1}_{X}(T_a,T_b,T_c)\medskip\\
(T_a,T_b,T_c)+\sigma^{-1}_{X'}(T_{2a},T_{2b},T_{2c})=(T_a,T_b,T_c)+\sigma^{-1}_{X'}(T_{1a},T_{1b},T_{1c})
\end{array}
\right.,
\end{equation*}
whose only solution is
$$(T_{1a},T_{1b},T_{1c})=(T_{2a},T_{2b},T_{2c}).$$ Therefore, $B_1$
and $B_2$ must be exactly the same. By the same token, this should
be true for the case of left exchange interaction too.

We now investigate the asymmetry of the second kind. We assume a
right-reducible braid
$B_1=\left.^{S_{1l}}\hspace{-0.5mm}[(T_{1a},T_{1b},T_{1c})\sigma_{X_1X'_1}]^{S}\right.$
with $X'_1$ the choice of its reducible crossing segment, and
$B_2=\left.^{S}\hspace{-0.5mm}[(T_{2a},T_{2b},T_{2c})\sigma_{X'_2X_2}]^{S_{2r}}\right.$
with $X'_2$ the choice of its reducible crossing segment. Note that
the right end-node of $B_1$ has already been made the same as the
left end-node of $B_2$ such that
$B_1\accentset{\rightarrow}{+}_{\mathrm{e}} B_2$ and
$B_1\accentset{\leftarrow}{+}_{\mathrm{e}} B_2$ are both allowed.

By Theorem \ref{theoType2Int} we get
\begin{align*}
&B_1\accentset{\rightarrow}{+}_{\mathrm{e}} B_2 \rightarrow\\
&\left.^{S_{1l}}\hspace{-0.5mm}[((T_{1a},T_{1b},T_{1c})-\sigma^{-1}_{X_1}(T_a,T_b,T_c))\sigma_{X_1}]^{S'}\right.
+
\left.^{S'}\hspace{-0.5mm}[((T_a,T_b,T_c)+\sigma^{-1}_{X'_1}(T_{2a},T_{2b},T_{2c}))\sigma_{X'_1X'_2X_2}]^{S_{2r}}\right.,
\end{align*}
where $S'=(-)^{|X'_1|}S$, and $(T_a,T_b,T_c)$ the triple of internal
twists of the virtual actively interacting braid in the interaction.
Analogously we also have
\begin{align*}
&B_1\accentset{\leftarrow}{+}_{\mathrm{e}} B_2 \rightarrow\\
&\left.^{S_{1l}}\hspace{-0.5mm}[((T_{1a},T_{1b},T_{1c})+\sigma^{-1}_{X_1X'_1X'_2}(T'_a,T'_b,T'_c))\sigma_{X_1X'_1X'_2}]^{S''}\right.
+
\left.^{S''}\hspace{-0.5mm}[((T_{2a},T_{2b},T_{2c})\sigma_{X'_2}-(T'_a,T'_b,T'_c))\sigma_{X_2}]^{S_{2r}}\right.,
\end{align*}
where $S''=(-)^{|X'_2|}S$, and $(T'_a,T'_b,T'_c)$ the triple of
internal twists of the virtual actively interacting braid in the
interaction. The term-by-term equality of the RHS of the above two
equations now helps us to pin down the conditions on $B_1$ and
$B_2$. Firstly, we must have $S'=S''$ which requires
$|X'_1|=|X'_2|+2n, n\in\mathbb{Z}$ and keeping $|X'_1|\geq 0$.
Secondly, we demand $X_1=X_1X'_1X'_2$ and $X'_1X'_2X_2=X_2$. The
only solution of this is readily $X'_1X'_2=\mathbb{I}$. Or
equivalently, we obtain $X'_1=X'^{-1}_2$. This sets $|X'_1|=|X'_2|$
and hence guarantees $S'=S''$. These results now turn the constraint
on the triples of internal twists of $B_1$ and $B_2$ to be
\begin{equation}
\begin{aligned}
&(T_{1a},T_{1b},T_{1c})-\sigma^{-1}_{X_1}(T_a,T_b,T_c)=
(T_{1a},T_{1b},T_{1c})+\sigma^{-1}_{X_1}(T'_a,T'_b,T'_c)\\
&(T_a,T_b,T_c)+\sigma^{-1}_{X'_1}(T_{2a},T_{2b},T_{2c})=
(T_{2a},T_{2b},T_{2c})\sigma_{X'_2}-(T'_a,T'_b,T'_c)
\end{aligned}.\label{eqTtriple2}
\end{equation}
Surprisingly, Eq. \ref{eqTtriple2} actually puts no more constraint
on $B_1$ and $B_2$ because it is automatically satisfied. The reason
is as follows.

In both cases, namely
$B_1\accentset{\rightarrow}{+}_{\mathrm{e}}B_2$ and
$B_1\accentset{\leftarrow}{+}_{\mathrm{e}}B_2$, the configurations
obtained by the first $2\rightarrow 3$ move are the same, which has
three new nodes and three new edges (like the dashed green ones in
Fig. \ref{type2IntDef}(b)). We call this configuration, $\Delta$. In
the case of $B_1\accentset{\rightarrow}{+}_{\mathrm{e}}B_2$, one
needs to translate the $\Delta$ to the left, passing through $X'_1$,
then rearrange $\Delta$ by equivalence moves into another
configuration, say $\Delta '$, which is a proper for a $3\rightarrow
2$ move. The equivalence moves taking $\Delta$ to $\Delta'$ equips
$\Delta'$ with two opposite triples of twists, viz $-(T_a, T_b,T_c)$
on the left of $\Delta'$ and $(T_a, T_b,T_c)$ on the right. $(T_a,
T_b,T_c)$ is the very triple of internal twists of the virtual
actively interacting braid exchanged by $B_1$ and $B_2$ in this
interaction. In the case of
$B_1\accentset{\leftarrow}{+}_{\mathrm{e}}B_2$, one translates
$\Delta$ to the right, passing the crossing sequence $X'_2$, and
reforms it by equivalence moves into another one, say $\Delta''$,
which is also ready for a $3\rightarrow 2$ move. Likewise,
$\Delta''$ is equipped with $(T'_a, T'_b,T'_c)$ on its left and
$-(T'_a, T'_b,T'_c)$ on its right. $(T'_a, T'_b,T'_c)$ is the very
triple of internal twists of the virtual actively interacting braid
exchanged in the process of
$B_1\accentset{\leftarrow}{+}_{\mathrm{e}}B_2$. Nevertheless,
because of $X'_2=X'^{-1}_1$, according to \cite{HeWan2008a},
$\Delta'$ and $\Delta''$ happen to be related to each other by a
discrete transformation which is considered as a $CP$
transformation. By the action property of a $CP$ (see
\cite{HeWan2008a} for details), we have precisely
\begin{equation}
(T'_a, T'_b,T'_c)=-(T_a, T_b,T_c).\label{eqCPtwist}
\end{equation}

Putting this aside first, we have another useful identity from
\cite{HeWan2008a}\footnote{The equation A.2 in the reference}:
$$\sigma^{-1}_X(\cdot,\cdot,\cdot)=(\cdot,\cdot,\cdot)\sigma_{\mathcal{R}(X)},$$
where $(\cdot,\cdot,\cdot)$ stands for an arbitrary triple of
internal twists, and $\mathcal{R}(X=x_1\dots x_i\dots x_n)=x_n\dots
x_i\dots x_1$, the reversion operation of a crossing sequence.
\cite{HeWan2008a} also defines the inversion of a crossing sequence,
namely $\mathcal{I}(X=x_1\dots x_i\dots x_n)=x^{-1}_1\dots
x^{-1}_i\dots x^{-1}_n$. Obviously,
$\mathcal{I}\mathcal{R}(X)=X^{-1}$. However, regarding the
permutation induced by a crossing sequence,
$\sigma_{\mathcal{R}(X)}=\sigma_{\mathcal{I}\mathcal{R}(X)}$\cite{HeWan2008a},
which then means $\sigma_{\mathcal{R}(X)}=\sigma_{X^{-1}}$.
Consequently, the equation above is extended to
\begin{equation}
\sigma^{-1}_X(\cdot,\cdot,\cdot)=(\cdot,\cdot,\cdot)\sigma_{X^{-1}}.\label{eqSigma}
\end{equation}

Finally, in view of Eqs. \ref{eqCPtwist} and \ref{eqSigma}, and all
the relations we have in hand, it is easy to see that Eq.
\ref{eqTtriple2} is satisfied. That is, the triples of internal
twists of $B_1$ and $B_2$ can be any thing they can. Therefore, for
$B_1\accentset{\rightarrow}{+}_{\mathrm{e}} B_2$ and
$B_1\accentset{\leftarrow}{+}_{\mathrm{e}} B_2$ can be equal in
certain way, we demand
$B_1=\left.^{S_{1l}}\hspace{-0.5mm}[(T_{1a},T_{1b},T_{1c})\sigma_{X_1X}]^{S}\right.$
and
$B_2=\left.^{S}\hspace{-0.5mm}[(T_{2a},T_{2b},T_{2c})\sigma_{X^{-1}X_2}]^{S_{2r}}\right.$.
For compactness, the discussion above is concluded by the following
theorem.
\begin{theorem}
Exchange interaction between two arbitrary braids
$B_1,B_2\in\mathfrak{B}^S$, if allowed, is asymmetric in general.
The asymmetry of the first kind goes like,
$B_1\accentset{\rightarrow}{+}_{\mathrm{e}} B_2\neq
B_2\accentset{\rightarrow}{+}_{\mathrm{e}} B_1$
($B_1\accentset{\leftarrow}{+}_{\mathrm{e}} B_2\neq
B_2\accentset{\leftarrow}{+}_{\mathrm{e}} B_1$), whereas the
asymmetry of the second kind reads
$B_1\accentset{\rightarrow}{+}_{\mathrm{e}} B_2\neq
B_1\accentset{\leftarrow}{+}_{\mathrm{e}} B_2$. However, there exist
the following special cases.
\begin{enumerate}
\item For that there exists a way in which $B_1\accentset{\rightarrow}{+}_{\mathrm{e}} B_2=
B_2\accentset{\rightarrow}{+}_{\mathrm{e}} B_1$
($B_1\accentset{\leftarrow}{+}_{\mathrm{e}} B_2=
B_2\accentset{\leftarrow}{+}_{\mathrm{e}} B_1$), $B_1$ and $B_2$
must be in the form
\begin{equation}
B_1=B_2=\left.^{S}\hspace{-0.5mm}[(\cdot,\cdot,\cdot)\sigma_{XX'}]^{S}\right.
\ \
(B_1=B_2=\left.^{S}\hspace{-0.5mm}[(\cdot,\cdot,\cdot)\sigma_{X'X}]^{S}\right.),
\end{equation}
where $(\cdot,\cdot,\cdot)$ represents an arbitrary triple of
internal twists, and $X'$ is the chosen reducible crossing segment.
\item For that there exists a way in which $B_1\accentset{\rightarrow}{+}_{\mathrm{e}} B_2=
B_1\accentset{\leftarrow}{+}_{\mathrm{e}} B_2$, $B_1$ and $B_2$ have
to be like
\begin{equation}
\begin{aligned}
&B_1=\left.^{S_{1l}}\hspace{-0.5mm}[(\cdot,\cdot,\cdot)\sigma_{X_1X}]^{S}\right.\\
&B_2=\left.^{S}\hspace{-0.5mm}[(\cdot,\cdot,\cdot)\sigma_{X^{-1}X_2}]^{S_{2r}}\right.
\end{aligned},
\end{equation}
where $X$ and $X^{-1}$ are the specified reducible crossing segments
of $B_1$ and $B_2$ respectively during the two interactions.
\end{enumerate}
\label{theoType2Asymm}
\end{theorem}

The existence of exchange interaction brings the braids and their
dynamics closer to Quantum Field Theory. This is not to say that we
already have a fundamental field theoretic formulation of our
theory. Rather, we can have an effective field theory describing the
dynamics of braids. To make this more transparent, we need first to
show another dynamical process of braids.

\section{Braid Decay}
It is implied in \cite{LeeWan2007} that there can be reversed
processes of direct interactions: a braid may split into two braids.
\begin{figure}[h]
\begin{center}
  \includegraphics[height=4.91in,width=2.81in]{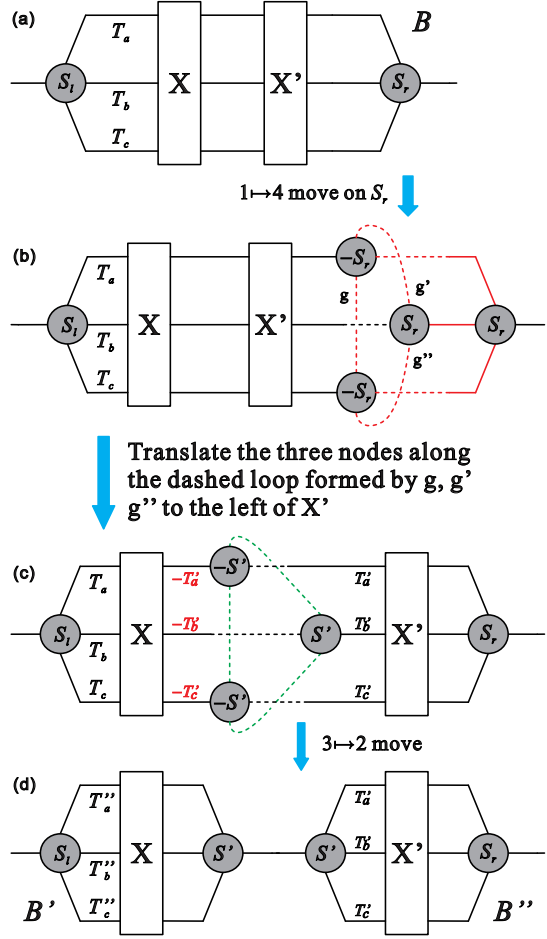}
  \caption{Definition of braid decay:
  $B\overset{\rightsquigarrow}{\rightarrow} B'+B''$,
  $B,B'\in\mathfrak{B}^S$, $B''\in\mathfrak{B}^b$. In (c) and (d), $S'=(-)^{|X'|}S_r$. In (b) and (c), the dashed lines
  emphasize the dependence of their relative positions
  on the state $S_r$. In (d), $(T''_a,T''_b,T''_c)$ is
  the triple of internal twists of $B'$.}
  \label{DecayDef}
\end{center}
\end{figure}
We may call a reversed direct interaction a decay. Hence, obviously
through decay a braid always radiates at least one actively
interacting braid. In a direct interaction, the actively interacting
braid involved may interact with the other braid from the left or
from the right. As a result, via a decay the braid may radiate an
actively interacting braid to its right or to its left, depending on
whether the braid is right-reducible or left-reducible. We thus
differentiate a left-decay from a right-decay. The former is
symbolized by $B\accentset{\leftsquigarrow}{\rightarrow}B'+B''$,
indicating that $B'\in\mathfrak{B}^b$, while the latter is denoted
by $B\accentset{\rightsquigarrow}{\rightarrow} B'+B''$ for that
$B''\in\mathfrak{B}^b$.

Fig. \ref{DecayDef} presents the defining process of right-decay.
Fig. \ref{DecayDef}(a) shows a right-reducible braid,
$B=\left.^{S_{l}}\hspace{-0.5mm}[(T_{a},T_{b},T_{c})\sigma_{XX'}]^{S_r}\right.$,
with a chosen reducible crossing segment $X'$. As in an exchange
interaction $X'$ can be a part of the maximal reducible crossing
segment of $B$'. This indicates that generically a braid can also
have different ways of decay, corresponding to each choice of the
reducible segment. We thus need to specify this choice in a specific
decay process.

Taking a $1\rightarrow 4$ move on the right end-node of $B$ leads to
Fig. \ref{DecayDef}(b). One then translate the three nodes, one in
state $S_r$ and two in $-S_r$, along the red dashed loop in Fig.
\ref{DecayDef}(b) to the left of $X'$, and rearrange them in a
proper configuration for a $3\rightarrow 2$ move, as shown in Fig.
\ref{DecayDef}(c). This produces the pair of opposite triples of
twists in Fig. \ref{DecayDef}(c), namely $-(T'_a,T'_b,T'_c)$ and
$(T'_a,T'_b,T'_c)$. Note that the triple $-(T'_a,T'_b,T'_c)$, which
is in red, should be understood to be added to $(T_a,T_b,T_c)$, the
original triple of internal twists of $B$, taking into account the
permutation induced by $X$, i.e.
$(T_a,T_b,T_c)-\sigma^{-1}_X(T'_a,T'_b,T'_c)=(T''_a,T''_b,T''_c)$.
Finally, a $3\rightarrow 2$ move results in Fig. \ref{DecayDef}(d)
which depicts the two resulted adjacent braids, $B'$ and $B''$. Thus
we have
\begin{equation}
\begin{aligned}
B=\left.^{S_{l}}\hspace{-0.5mm}[(T_{a},T_{b},T_{c})\sigma_{XX'}]^{S_r}\right.
&\accentset{\rightsquigarrow}{\rightarrow}
\left.^{S_{l}}\hspace{-0.5mm}[((T_a,T_b,T_c)-\sigma^{-1}_X(T'_a,T'_b,T'_c))\sigma_X]^{S'}\right.+
\left.^{S'}\hspace{-0.5mm}[(T'_a,T'_b,T'_c)\sigma_{X'}]^{S_r}\right.\\
&=B'+B''
\end{aligned},\label{eqDecay}
\end{equation}
where $S'=(-)^{|X'|}S_r$ and $B''\in\mathfrak{B}^b$. Left-decay is
defined similarly.

By the same logic as that of the remark below Theorem
\ref{theoType2Int}, an irreducible braid can also decay but it only
radiates a completely trivial braid which is either $B_0^+$ or
$B_0^-$ in Eq. \ref{eqB0pm}. An irreducible braid hence stays
unchanged topologically under a decay.

The notation of decay already implies that a decay is in general not
symmetric. For a braid $B\in\mathfrak{B}^f\sqcup\mathfrak{B}^b$
which is able to decay in both directions, say
$B\accentset{\leftsquigarrow}{\rightarrow}B_1+B_2$ and
$B\accentset{\rightsquigarrow}{\rightarrow}B'_1+B'_2$, there is no
way for $B_1=B'_1$ and $B_2=B'_2$ because we must have
$B_1,B'_2\in\mathfrak{B}^b$ but
$B_2,B'_1\in\mathfrak{B}\setminus\mathfrak{B}^b$. Even if $B$ is an
actively interacting braid, its left and right decays give rise to
different results because both of its left and right decays have
more than one ways to occur, as pointed out above. There are also
special cases in which an actively interacting braid has a certain
left decay that is the same as a certain right decay of it. The
conditions of this are similar to those for a direct interaction to
be symmetric found in \cite{HeWan2008b}.

On the other hand, a direct interaction is not guaranteed to have a
corresponding decay despite that a decay can be viewed as the
reverse of certain direct interaction. This is because pair
cancelation of crossings may occur in a direct interaction and in
\cite{HeWan2008b} it is pointed out that in the definition of a
3-strand braid the crossing sequence of the braid in any
representation is the shortest one among all equivalent ones due to
braid relations. Indeed, one can easily show that a braid in
$\mathfrak{B}^s$ cannot decay into a braid in $\mathfrak{B}^f$,
although a direct interaction in the opposite direction is always
possible.

Because of the relation between decay and direct interaction,
effective twist $\Theta$ and effective state $\chi$ must be
respectively an additive conserved quantity and a multiplicative
conserved quantity under a decay. By the same token, because direct
interactions are invariant under C, P, T, and their
combinations\cite{HeWan2008a}, so are decays. Decay and direct
interaction of braids indicate that an actively interacting braid
can be singly created and destroyed. This reinforces the implication
that actively interacting braids are analogous to bosons.

\section{Braid Feynman diagrams}
Our study of braid excitations of embedded framed spin networks, in
particular the discovery of the dynamics of these excitations,
namely direct and exchange interactions, and decay of braids, makes
it possible to describe the dynamics of braids by an effective
theory based on Feynman diagrams. These diagrams are called braid
Feynman diagrams. We remark that as a distinction from the usual QFT
Feynman diagrams which do not have any internal structure, each
braid Feynman diagram is an effective description of the whole
dynamical process of an interaction of braids, its internal
structure records how the braids and their neighborhood evolve.

We use {\includegraphics[height=0.1219in,
width=0.6417in]{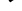}} and {\includegraphics[height=0.1219in,
width=0.6417in]{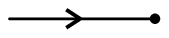}} for respectively outgoing and ingoing
propagating braids in $\mathfrak{B}^f$,
{\includegraphics[height=0.1219in, width=0.6417in]{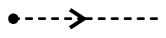}} and
{\includegraphics[height=0.1219in, width=0.6417in]{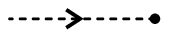}} for
respectively outgoing and ingoing stationary
braids\footnote{Stationary braids although are not directly
propagating on the spin network it resides, may still be able to
move around under the change of its surroundings due to the network
evolution, and may also propagate physically in the continuum limit
of the theory.} in $\mathfrak{B}^s$. Because it is implied that
actively interacting braids are analogous to bosons, outgoing and
ingoing braids in $\mathfrak{B}^b$ are better represented by
{\includegraphics[height=0.1643in, width=0.5025in]{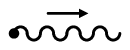}} and
{\includegraphics[height=0.1643in, width=0.5025in]{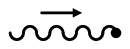}}
respectively.

In accordance with left and right decay, we will henceforth denote
left and right direct interactions by
$\accentset{\leftarrow}{+}_{\mathrm{d}}$ and
$\accentset{\rightarrow}{+}_{\mathrm{d}}$ respectively. Note that if
the two braids being interacting are both actively interacting, the
direction of the direct interaction is irrelevant because the result
does not depend on which of the two braids plays the actual active
role in the interaction\cite{LeeWan2007, HackettWan2008}. According
to the algebraic structure of braids under direct interaction found
in \cite{HeWan2008b}\footnote{See Theorem 4 in the reference.},
namely
$\mathfrak{B}^b\accentset{\rightarrow}{+}_{\mathrm{d}}\mathfrak{B}^b\subseteq\mathfrak{B}^b$
and
$\mathfrak{B}^b\accentset{\rightarrow}{+}_{\mathrm{d}}(\mathfrak{B}^f\sqcup\mathfrak{B}^s)\subseteq\mathfrak{B}^f\sqcup\mathfrak{B}^s$,
the only possible single vertices of right direct interactions are
listed below. (Those corresponding to left direct interactions are
left-right mirror images these.)
\begin{figure}[h]
\begin{center}
\includegraphics[height=0.7in,width=4.52in]{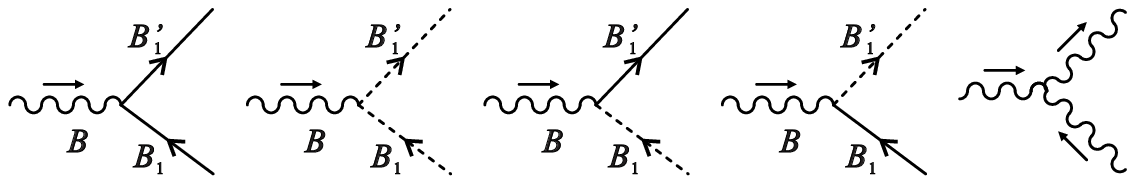}
\caption{Possible right direct interaction vertices. Time flows up.}
\label{rType1Feynman}
\end{center}
\end{figure}
\begin{figure}[h]
\begin{center}
\includegraphics[height=0.7in,width=3.72in]{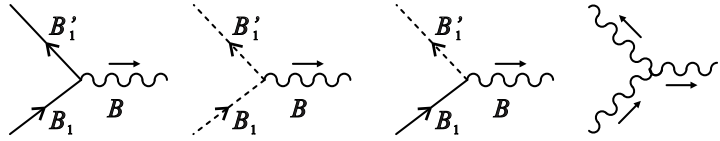}
\caption{Possible right-decay interaction vertices. Time flows up.}
\label{rDecayFeynman}
\end{center}
\end{figure}

Similarly, right decay have the possible basic single vertices in
Fig. \ref{rDecayFeynman}. (The left-decay vertices are left-right
mirror images of these.) In the second figure of Fig.
\ref{rDecayFeynman} when $B_1\in\mathfrak{B}^s$ is irreducible,
$B_1$ should be understood equal to $B'_1$ and $B$ is either of the
two braids in Eq. \ref{eqB0pm}.

The arrows over the wavy lines in Figs. \ref{rType1Feynman} and
\ref{rDecayFeynman} are important for that they differentiate a left
process from a right process. More importantly, they encode the fact
that the correspondence between decay and direct interaction is not
one-to-one and can be used to see which kind of direct interactions
can have a corresponding decay. That said, one can flip the arrow
over the braid $B$ in each of the first three diagrams in Fig.
\ref{rDecayFeynman} and obtain the diagram of the corresponding left
direction interaction. One can also turn the fourth diagram in Fig.
\ref{rDecayFeynman} upside down to get its corresponding direct
interaction diagram. However, for example, one cannot flip the arrow
over the wavy line in the fourth diagram in Fig. \ref{rType1Feynman}
to obtain a decay of the braid $B_1$ in the diagram because there
does not exist such a left decay according to the mirror images of
the diagrams in Fig. \ref{rDecayFeynman}.

\begin{figure}[h]
\begin{center}
\includegraphics[height=4.12in,width=6.30in]{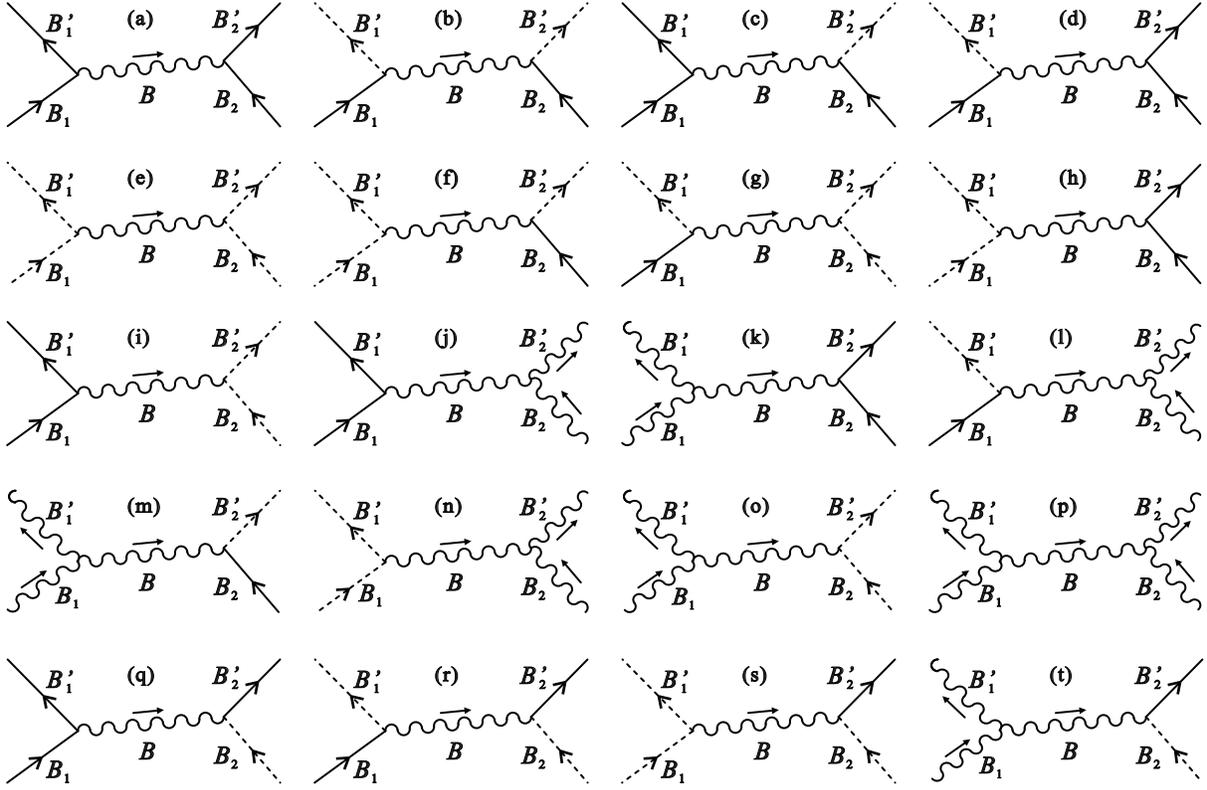}
\caption{Possible right exchange interaction 2-vertices. Time flows
up.} \label{rType2Feynman}
\end{center}
\end{figure}
The result of an exchange interaction of two braids, say
$B_1\accentset{\rightarrow}{+}_{\mathrm{e}}B_2\rightarrow B'_1+B'_2$
by exchanging some virtual $B\in\mathfrak{B}^b$, is the same as that
of the combined process of the decay,
$B_1\accentset{\rightsquigarrow}{\rightarrow}B'_1+B$, and the direct
interaction $B\accentset{\rightarrow}{+}_{\mathrm{d}}B_2\rightarrow
B'_2$. However, it is important to note that the former process and
the latter combined process are two topologically and dynamically
distinct processes; they only have the same in and out states
topologically. Therefore, we obtain in Fig. \ref{rType2Feynman} all
possible basic 2-vertex diagrams for right exchange interaction.

The left-right mirror images of the diagrams in Fig.
\ref{rType2Feynman} are certainly the basic diagrams for left
exchange interaction. In Fig. \ref{rType2Feynman}(e), (f), (h), (n),
(s) and their left-right mirror images, if $B_1\in\mathfrak{B}^s$
happens to be irreducible, $B_1=B'_1$ and $B=B_0^+$ or $B=B_0^-$ are
understood. Given all these it becomes manifest that exchange
interaction is invariant under the C, P, T and their products
defined in \cite{HeWan2008a}, as in the case of direct interaction
and decay.

Recalling Theorem \ref{theoType2Asymm}, braid Feynman diagrams make
it transparent to tell whether an exchange interaction can be
symmetric. Regarding the asymmetry of the first kind, one simply
needs to check if the diagram of an interaction looks the same as
its left-right mirror with also the arrow over the virtual braid
reversed. Fig. \ref{rType2Feynman}(c), (d), (h) through (o), and (q)
through (t) depict exchange interactions which do not allow any
violation of the asymmetry of the first kind, viz
$B_1\accentset{\rightarrow}{+}_{\mathrm{e}} B_2\neq
B_2\accentset{\rightarrow}{+}_{\mathrm{e}} B_1$. The reason is that
in each of these diagrams, the out-state contains two braids in two
different divisions respectively. Their mirror images have the same
kind of asymmetry with respect to left exchange interaction.

As to the asymmetry of the second kind, one should check if the
arrow over the virtual braid in a diagram can be flipped. According
to this, the four diagrams in Fig. \ref{rType2Feynman}(q) through
(t) respect the asymmetry of the second kind, i.e.
$B_1\accentset{\rightarrow}{+}_{\mathrm{e}} B_2\neq
B_1\accentset{\leftarrow}{+}_{\mathrm{e}} B_2$ in any case, because
the arrow over the braid $B$ in any of these diagrams cannot be
flipped, for which the fact that a stationary braid does not decay
into a propagating braid is accounted. Therefore, the exchange
interactions respectively in Fig. \ref{rType2Feynman}(q) to (t) are
completely asymmetric.

Nevertheless, interactions respectively in Fig.
\ref{rType2Feynman}(a) through (p) can have instances violating the
asymmetry of the second kind if conditions in Theorem
\ref{theoType2Asymm} are satisfied, because these diagrams are
symmetric under flipping the arrow over the braid $B$ in each of
them. As a result, the exchange interactions which can be fully
symmetric when conditions in Theorem \ref{theoType2Asymm} are
satisfied correspond respectively to the diagrams in Fig.
\ref{rType2Feynman}(a), (b), (e), (f), (g), and (p).

The analogy between actively interacting braids and bosons is
manifested by the braid Feynman diagrams in Figs.
\ref{rType1Feynman}, \ref{rDecayFeynman}, and \ref{rType2Feynman}.
Fermionic degrees of freedom may correspond to those braids which
are not actively interacting because their interactions are mediated
by the actively interacting ones. They are more probably
corresponding to braids in $\mathfrak{B}^f$, which are chiral
propagating but not actively interacting.

More complicated braid Feynman diagrams including the loop ones can
be constructed out of these basic vertices. As a result, there
should exist an effective field theory based on these diagrams, in
which the probability amplitudes of each diagram can be computed.
For this one should figure out the terms evaluating external lines,
vertices, and propagators of braids.

In a more complete sense, one may try to write down an action of the
effective fields representing braids that can generate these braid
Feynman diagrams. In such an effective theory, each line of a braid
Feynman diagram represents an effective field characterizing a
braid; it should be labeled by characterizing quantities of the
braid represented by the line, which are elements of certain groups
or the corresponding representations of these groups. For a braid
these quantities are its two end-node states which are elements of
$\mathbb{Z}_2$, its crossing sequence, an element of the braid group
$B_3$, its twists which are elements in $\mathbb{Z}$, and spin
network labels when they are taken into account. Moreover, there are
constraints of these group elements on the lines meeting at a
vertex.

This is more difficult than just to find a way to compute the
probability amplitude of each braid Feynman diagram. In any case,
the very first challenge of fulfilling this task is to choose an
appropriate mathematical language. In the next section we will
briefly mention three possible formalisms.

However, these braid Feynman diagrams put a constraint on defining
the probability amplitudes regardless of the underlining
mathematical language. We use an example to illustrate this point.
Let us consider the braid Feynman diagram in Fig.
\ref{rType2Feynman}(a) for some specific interaction. This diagram
is the same as concatenating the first diagram in Fig.
\ref{rDecayFeynman} with the first diagram in Fig.
\ref{rType1Feynman} from the left along the wavy line. That is, the
exchange interaction,
$B_1\accentset{\rightarrow}{+}_{\mathrm{e}}B_2\rightarrow B'_1+B'_2$
by exchanging some virtual $B\in\mathfrak{B}^b$, and the combined
process of the decay,
$B_1\accentset{\rightsquigarrow}{\rightarrow}B'_1+B$, and the direct
interaction $B\accentset{\rightarrow}{+}_{\mathrm{d}}B_2\rightarrow
B'_2$, have identical in and out braid states at this tree level. We
hence expect the following equality at this level, which is only
formal,
\begin{equation}
\mathcal{M}\left(B_1\accentset{\rightarrow}{+}_{\mathrm{e}}B_2\rightarrow
B'_1+B'_2\right) =
\sum\limits_{\alpha}\mathcal{M}\left(B\accentset{\rightarrow}{+}_{\mathrm{d}}B_2\rightarrow
B'_2\left|B_1\accentset{\rightsquigarrow}{\rightarrow}B'_1+B\right.,
\alpha\right)G_B(\alpha). %
\label{eqM}
\end{equation}

The LHS of Eq. \ref{eqM} is the probability amplitude of the
exchange interaction, which is independent of the virtual braid $B$
being exchanged during the interaction but only determined by the
evolution moves involved in the interaction and the external lines
in Fig. \ref{rType2Feynman}(a), namely $B_1$, $B_2$, $B'_1$ and
$B'_2$. The first term on the RHS of this equation is the
conditional probability amplitude of the direct interaction provided
with the occurrence of the decay and the meeting of $B$ and $B_2$.
Besides $B_1$, $B_2$, $B'_1$, $B'_2$ and the corresponding evolution
moves, this also certainly depends on the braid $B$ which is
characterized by a set of parameters, denoted by $\alpha$, including
the end-node states, spin network labels, twists, and crossings of
$B$. The second term on the RHS, i.e. $G_B(\alpha)$ represents the
propagator (however it will be defined) of $B$, which is obviously
also a function of $\alpha$. $\alpha$ must be summed over due to the
independence of the final result on $B$. There are analogies of this
summation in usual quantum field theories, e.g. the integration over
the momentum defining the propagator of the virtual particle in an
interaction, and the summation over polarizations of gauge bosons.

Eq. \ref{eqM} is generic though is derived with the help of a
specific example because one can simply replace the exchange
interaction on the LHS with any other one and substitute the
corresponding direct interaction and decay on the RHS
simultaneously. No matter how the continuum limit of our theory is
to be obtained, the momentum of the braid $B$ in this limit should
be accounted for by $\alpha$. Therefore, Eq. \ref{eqM} can be used
for a validity check of the theory's future possible developments
which will be able to define the probability amplitudes of the
dynamical processes of our braids.

\section{Conclusions and future work}
In conclusion, we have found the exchange interaction of braids,
which has two kinds of asymmetry. Conserved quantities under
exchange interaction are discussed. We also discussed decay of
braids. The existence of exchange interaction and its relation with
direct interaction and decay of braids imply the analogy between
actively interacting braids and bosons. Braid Feynman diagrams are
developed and used to represent the dynamics of braids. An effective
theory describing braid dynamics can be based on these braid Feynman
diagrams. We emphasize that an interaction of two braids is not
point-like, although braid Feynman vertices are point-like. This is
similar to the case of String Theory in which two strings do no
interact at a point.

Despite the lack of a fully fundamental theory of quantum gravity
with matter, an effective theory of topological excitations, such as
our braids, of quantum geometry may be more relevant to the testable
region of our physical world. The study of collective modes in
condensed matter physics provides a great motivation to this. For
example, in the current stage of the string-net condensation, all
Standard Model gauge fields and fermionic fields but chiral fermions
appear to be low energy effective fields emergent out of certain
high energy lattice models\cite{WenXiaogang}.

Our next step is to compute the probability amplitudes of the braid
Feynman diagrams and to write down the effective field theory of
these braid excitations in an algebraic way. To compute the
probability amplitudes, one may adopt the methods in some spin foam
models, in which the probability amplitude of each evolution move is
most basic. However, the way to define the probability amplitude of
an evolution move in our situation should be modified because all
current spin foam models are unembedded but our spin networks are
embedded in a topological 3-manifold, which means that not only spin
network labels but also the topological quantities defining the
braids should be considered.

One may also borrow the ideas from Group Field Theories in which a
field is a scalar function of group elements defining a fundamental
building block of spacetime\cite{Oriti2007}. We can try to construct
a group field theory of braids, in which a field is a function of
the group elements defining a braid. The interacting term in this
theory can be written done with the help of braid Feynman diagrams.
The very first challenge is to construct the field representing a
braid. As a first step, one may start with a toy model, considering
actively interacting braids only, or even merely certain actively
interacting braids, of a spin network. This is what we are currently
working on.

The third possibility is by means of tensor categories. The
motivations of adopting tensor categorical methods have been
discussed in \cite{HeWan2008a}. The connection between LQG and Spin
Foam Models and Tensor Categories has actually been realized for
about two decades. It was first introduced by Crane\cite{Crane1991}
and elaborated by others, e.g. Kauffman\cite{Kauffman}. One should
also note that the string network condensation by Wen \emph{et
al}\cite{WenXiaogang} and their newly proposed tensor-net
approach\cite{Wen2008} are examples of approaches of unifying
gravity and matter which indicate that tensor category might be a
correct underlying mathematical language towards this goal. This
formulation would be purely algebraically, referring to any
topological embedding is no longer needed.

Another open issue is that we cannot justify for now whether the
actively interacting braids are analogous to bosons or gauge bosons
in particular. For the latter to be true, actively interacting
braids should obey certain gauge symmetry. We would like to see
gauge symmetries arise when we include spin network labels, which
are normally gauge group representations, in our model. The three
possible approaches introduced above may have an answer for this
question.

\section*{Acknowledgements}
The author thanks Sundance Bilson-Thompson, Jonathan Hackett, and
Matthias Wapler for helpful discussions. His gratitude also goes to
Daniele Oriti and Florian Conrady for clarifying some GFT concepts.
He appreciates Zhengcheng Gu for discussions on tensor categories
and the clarification of Wen's work. He is grateful to Song He and
Lee Smolin, his advisor, for invaluable comments on the manuscript.
Research at Perimeter Institute for Theoretical Physics is supported
in part by the Government of Canada through NSERC and by the
Province of Ontario through MRI.

\end{document}